\documentclass[aps,prl,superscriptaddress,amsmath,amssymb,twocolumn,nofootinbib]{revtex4}

\usepackage[utf8]{inputenc}
\usepackage{color}
\usepackage{graphicx}
\usepackage{dashbox}
\usepackage{physics}

\def\JGB#1{#1}
\def\half{\frac{1}{2}}

\def\OM{\Omega_{\rm M}}
\def\OK{\Omega_{\rm K}}
\def\OL{\Omega_\Lambda}

\begin{document}

\thispagestyle{empty}

\title{Dark Energy predictions from GREA: Background and linear perturbation theory}

\author{Juan Garc\'ia-Bellido}\email[]{juan.garciabellido@uam.es}
\affiliation{Instituto de F\'isica Te\'orica UAM-CSIC, Universidad Auton\'oma de Madrid, Cantoblanco 28049 Madrid, Spain}

\date{\today}

\preprint{IFT-UAM/CSIC-24-69}

\begin{abstract}

General Relativistic Entropic Acceleration (GREA) theory provides a covariant formalism for out-of-equilibrium phenomena in GR, extending the Einstein equations with an entropic force that behaves like bulk viscosity with a negative effective pressure. In particular, the growth of entropy associated with the homogeneous causal horizon can explain the present acceleration of the Universe, without introducing a cosmological constant. The dynamics of the accelerated universe is characterized by a single parameter $\alpha$, the ratio of the causal horizon to the curvature scale, which provides a unique history of the Universe distinguishable from that of $\Lambda$CDM. In particular, we explain the coincidence problem and the Hubble tension by shifting the coasting point to higher redshifts. All background observables are correlated among themselves due to their common dependence on $\alpha$. This scenario gives a specific evolution for the effective equation of state parameter, $w(a)$.
Furthermore, we study the linear growth of matter perturbations in the context of a homogeneous expanding background driven by the entropy of the causal horizon. We find that the rate of growth of matter fluctuations in GREA slows down due to the accelerated expansion and alleviates the $\sigma_8$ tension of $\Lambda$CDM. We compute the growth function of matter fluctuations, the redshift space distortions in the galaxy correlation function, as well as the redshift evolution of the baryon acoustic oscillation scale, and find that the integrated Sachs-Wolfe effect is significantly larger than in $\Lambda$CDM. It is interesting to note that many of the tensions and anomalies of the standard model of cosmology are alleviated by the inclusion of this transient period of acceleration of the Universe based on known fundamental physics. In the near future we will be able to constrain this theory with present data from deep galaxy surveys.

\end{abstract}
\maketitle

\section{I. Introduction}

General Relativistic Entropic Acceleration (GREA) theory~\cite{Espinosa-Portales:2021cac} gives a covariant formalism for out of equilibrium dynamics in the context of general relativity. 
A consequence of GREA is the explicit breaking of the time reversal invariance when there is entropy production. This drives an entropic force that behaves effectively like bulk viscosity with a negative effective pressure. The natural consequence of this formalism is that the source for space-time curvature is the Helmholtz free energy, $F = U - TS$, and not just matter and radiation~\cite{Garcia-Bellido:2024tip,Garcia-Bellido:2023wcw}. 

When the entropy production is associated with the growth of horizons one can describe their thermodynamical effects as a GHY boundary term~\cite{Gibbons:1976ue,York:1972sj}. In the cosmological context, this theory predicts that the present acceleration of the universe could arise from the growth of entropy associated with cosmic~\cite{Garcia-Bellido:2021idr} and black hole horizons~\cite{Garcia-Bellido:2024tip}, without the need to introduce a cosmological constant, providing an alternative dynamics to that of the Standard Cosmological Model ($\Lambda$CDM). 

Already before the recent evidence of significant deviations from the standard model of cosmology through state-of-the-art observations by DES-Y5 SNe~\cite{DES:2024tys} and DESI-Y1 BAO~\cite{DESI:2024mwx}, GREA predictions gave a better  description of cosmological observations than $\Lambda$CDM~\cite{Arjona:2021uxs}.
In particular, GREA explains the coincidence problem and can alleviate the Hubble tension by shifting the coasting point (when the universe transitioned from matter domination to acceleration) to higher redshifts and extending the period of acceleration. 

In this paper, we perform a detailed calculation of the background and linear matter perturbations with the assumption that the entropy growth responsible for the accelerated expansion comes from the {\em homogeneous} cosmological horizon.
We  show that, in this case, the dynamics can be described in terms of just one parameter, $\alpha$, the ratio of the spatial curvature to the horizon distance, which gives the size of the cosmological horizon and the rate of growth of the horizon entropy. A given value of $\alpha$ defines a unique history of the universe, with a dynamics that differs from $\Lambda$CDM and gives an explicit prediction for all background observables, like the values of the matter and dark energy content of the universe, the rate of expansion today, the coasting point, the value of the equation of state of dark energy and its derivative, the angular diameter and luminosity distances, etc. 

In order to compare with the plethora of cosmological data, we need not just the background evolution but also the linear growth of matter fluctuations. We derive a second order differential equation in GREA for the amplitude of matter perturbations in a homogeneous background, and compare with the corresponding one in $\Lambda$CDM. The solutions match those during the matter era, but start deviating around redshift one, when the entropic forces begin to take over and start accelerating the universe.

In Section II we describe the extended Einstein field equations in the presence of out-of-equilibrium phenomena associated with the growth of horizons. We recall from Ref.~\cite{Espinosa-Portales:2021cac} that the first law of thermodynamics can be incorporated into Einstein's equations as an effective bulk viscosity, which drives an effective {\em negative} pressure as a consequence of the second law of thermodynamics.  In Section III we explore the background phenomenology of GREA in the context of entropic forces arising from the cosmological horizon. In Section IV we derive the linear perturbation theory of GREA and compare cosmological observables with the corresponding ones in $\Lambda$CDM. Finally, we conclude in Section V. 

\section{II. The Einstein field equations in GREA}

We give here a short summary of the covariant formalism developed in Ref.~\cite{Espinosa-Portales:2021cac} that describes the extended Einstein field equations in the presence of out-of-equilibrium phenomenon associated with the growth of horizons.
The gravitational action of the Standard Model of Cosmology ($\Lambda$CDM) is given by
$${\cal S} = \frac{1}{2\kappa}\int_{\cal M}\!d^4x\sqrt{-g}\Big[R+2\kappa{\cal L}_m\Big] - \frac{1}{\kappa}
\int_{\cal M}\!d^4x\sqrt{-g}\,\Lambda\,,$$
where ${\cal L}_m$ is the matter Lagrangian, $\kappa = 8\pi G$ and $\Lambda$ is the cosmological constant. 
In the context of GREA we substitute the bulk term of the cosmological constant for the GHY boundary term of the horizon~\cite{Garcia-Bellido:2021idr},
$${\cal S} = \frac{1}{2\kappa}\int_{\cal M}\!d^4x\sqrt{-g}\Big[R+2\kappa{\cal L}_m\Big]  + \frac{1}{\kappa}
\int_{\cal H}\!d^3x\sqrt{h}\,K\,,$$
where $K$ is the extrinsic curvature and $h$ the 3D metric of the cosmological horizon, ${\cal H}=\partial{\cal M}$.

The main difference of GREA with respect to $\Lambda$CDM is the breaking of time-reversal invariance at the level of the action by the entropic force term that arises due to the growth of the cosmic horizon. This introduces a cosmological arrow of time. Most of the evolution of the universe has been quasi-adiabatic since these entropic forces acted only during inflation, reheating after inflation and at the present time, when the universe is old enough and the cosmic horizon associated with radiation and matter has grown sufficiently large for its entropy growth to dominate over the attraction of matter and induce the present acceleration. Eventually, even this period of acceleration will end as the entropic term gets diluted and we will end in an empty Minkowsky space-time~\cite{Garcia-Bellido:2021idr}. 

Therefore, the Einstein field equations, 
{\em extended} to out-of-equilibrium phenomena, can be written as~\cite{Espinosa-Portales:2021cac},
\begin{equation}
	\label{eq:GREA}
	G_{\mu\nu} = R_{\mu\nu} - \half R\,g_{\mu\nu} = \kappa\, (T_{\mu\nu} - f_{\mu\nu}) \equiv \kappa\,{\cal T}_{\mu\nu}\,,
\end{equation}
where $f_{\mu\nu}$ arises from the first law of thermodynamics, and introduces an effective {\em negative}  pressure associated with the growth of entropy, $p_S = - TdS/dV < 0$, according to the second law of thermodynamics. This extra component to the Einstein equations can be interpreted as an effective bulk viscosity term of a real (non-ideal) fluid~\cite{Espinosa-Portales:2021cac,Gagnon:2011id}, with $\Theta = D_\lambda u^\lambda$ the trace of the congruence of geodesics,
\begin{equation}
	\label{eq:fmunu}
	f_{\mu\nu} = \zeta\,\Theta \,(g_{\mu\nu}+u_\mu u_\nu) = \zeta\,\Theta\,h_{\mu\nu} \,,
\end{equation}
such that the covariantly-conserved energy-momentum tensor has the form of a perfect fluid tensor,
\begin{eqnarray}
	{\cal T}^{\mu\nu} &=& p\,g^{\mu\nu} + (\rho + p)u^\mu u^\nu -  \zeta\,\Theta\,h^{\mu\nu} \nonumber \\ &=& \tilde p\,g^{\mu\nu} + (\rho + \tilde p)u^\mu u^\nu\,,
    \label{eq:Tmunu}
\end{eqnarray}
with $\tilde p = p + p_S$, and the bulk viscosity coefficient $\zeta$ can be written as~\cite{Espinosa-Portales:2021cac}
\begin{equation}
	\label{eq:zeta}
	\zeta = \frac{T}{\Theta}\frac{dS}{dV}\,.
\end{equation}
In the case of an expanding universe, $\Theta=\frac{d}{dt}\ln V = 3H$ and the coefficient becomes $\zeta = T\dot S/(9H^2a^3)$, see~\cite{Garcia-Bellido:2021idr}, with $S$ the entropy per comoving volume of the Universe. Entropy production therefore implies $\zeta > 0$.

Note that the matter contribution to the Hamiltonian constraint arises from variations of the matter action with respect to the lapse function (and afterwards setting $N(t)=c$),
\begin{equation}
	\label{eq:matter}
	{\cal S}_m = \int\!d^4x\sqrt{-g}\,{\cal L}_m = - \int dt
    \,N(t)\,V_c\,a^3\rho\,,
\end{equation}
where $V_c$ is the {\em comoving} volume,
\begin{eqnarray}
	\label{eq:Vc}
    &&V_c = \int\!\frac{4\pi\,r^2\,dr}{\sqrt{1-k\,r^2}} \,=\, \frac{2\pi}{(-k)^{3/2}}\times\\
    &&\left[\ln\left(\sqrt{1-k\,r^2}-\sqrt{-k\,r^2}\right) + \sqrt{-k\,r^2(1-k\,r^2)}\right]\,, \nonumber
\end{eqnarray}
for a given radial coordinate $r$. The conservation of the total energy momentum tensor is derived from the first law of thermodynamics,
\begin{eqnarray}
	\label{eq:TdS}
	&&dU + pdV - TdS = 0 \hspace{3mm} \Rightarrow \\[3mm] &&V_c \left[ d(\rho\,a^3) +
    p\,d(a^3) - T\,dS_c\right] = 0 \hspace{3mm} \Rightarrow\\
    &&\dot\rho + 3H(\rho + p) = \frac{T\dot S_c}{a^3}\,,
\end{eqnarray}
where $S_c$  is the entropy per {\em comoving} volume. For adiabatic expansion, or constant $S_c$, we recover the usual conservation equation~\cite{Espinosa-Portales:2021cac}.

\subsection{The causal cosmological horizon}

Let us describe here the entropic forces induced by the causal cosmological horizon of a FLRW universe~\cite{Garcia-Bellido:2021idr}. We start by considering an arbitrary comoving 2-sphere around the origin of coordinates. Then the trace of its extrinsic curvature is given by (we set $c=1$ everywhere)
\begin{equation}\label{eq:Ext}
    \sqrt{h} K = - 2 N(t) \,r\, a \sqrt{1 - kr^2} \sin \theta   
\end{equation}
where $r=\sinh(\eta\sqrt{-k})/\sqrt{-k}$ along the lightcone, and $\eta$ is the conformal time. The boundary term for the causal cosmological horizon, $d_H = a\,\eta$, can be written as~\cite{Garcia-Bellido:2021idr} 
\begin{eqnarray}
    S_{\rm GHY} &=& - \frac{1}{2G} \int dt \,N(t)\,\frac{a}{\sqrt{-k}}\sinh(2\eta\sqrt{-k}) \label{eq:GHY} \\
    &=& - \int dt \,N(t)\,T_H S_H \ 
    \equiv - \int dt \,N(t)\,V_c\,a^3\rho_H\,, \nonumber
\end{eqnarray}
where $T_H$ is the temperature and $S_H$ the entropy associated with the causal cosmological horizon~\cite{Garcia-Bellido:2021idr}
\begin{equation}
    k_{\rm B}T_H = \frac{\hbar}{2\pi}\ 
    \frac{a\,\sinh(2\eta\sqrt{-k})}
    {d_H^2\sqrt{-k}}\,, \hspace{5mm} 
    S_H = \frac{k_{\rm B}\pi}{\hbar} \frac{d_H^2}{G}\,.
\end{equation}
The fact that we can naturally assign a temperature and an entropy to a hypersurface is a signal of the existence of an underlying quantum description of gravity and thermodynamics~\cite{Hawking:1974sw}. This is made explicit by the appearance of $\hbar$ in both quantities. Their product, however, does not depend on $\hbar$ and leads to a classical {\em emergent} phenomenon, the acceleration of the universe. Note also that variation w.r.t. the lapse function will give an extra contribution to the Hamiltonian constraint coming from the boundary term~\cite{Gibbons:1976ue}, whose origin is related to the quantum degrees of freedom of the horizon, a phenomenon which has connections with the Holographic principle~\cite{tHooft:1993dmi,Susskind:1994vu}.

\begin{figure}[t]
    \centering
    \includegraphics[width=\linewidth]{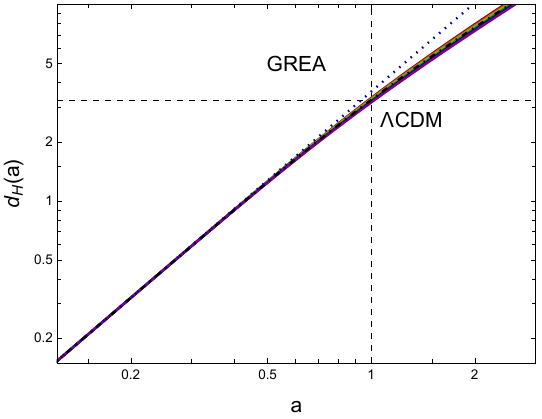}
    \caption{The horizon distance as a function of the scale factor in GREA (color lines), compared with $\Lambda$CDM (black dot-dashed line) and matter dominated (dotted blue line). It is clear that the ratio of the curvature scale to the horizon distance is very approximately one, for all realizations of GREA.}
    \label{fig:dH}
\end{figure}

\begin{figure}[t]
    \centering
    \includegraphics[width=0.97\linewidth]{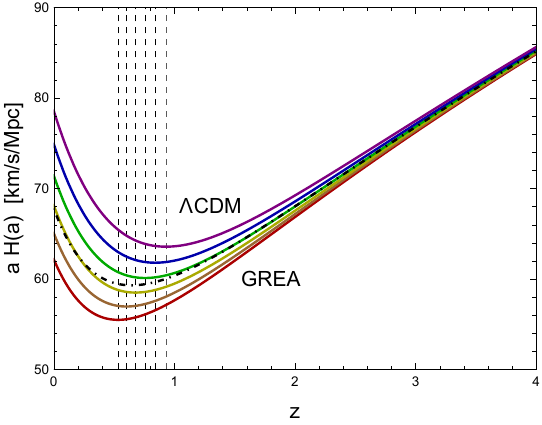}\\
    \vspace{5mm}
    \includegraphics[width=\linewidth]{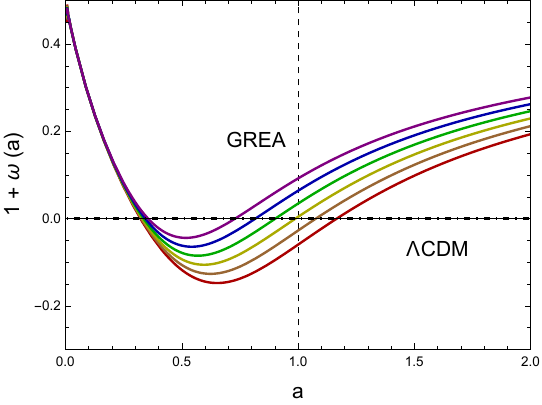}
    \caption{The upper panel shows the evolution of the inverse comoving horizon with the corresponding coasting points (dashed lines) for each value of $\alpha$ (in color). The lower panel shows the evolution of the effective DE equation of state $w(a)$.}
    \label{fig:CoastingPoint}
\end{figure}

\section{III. GREA phenomenology}

We can then solve the dynamical Friedmann equations for the accelerated universe in the context of an entropic force arising from the causal horizon. We will assume here a concrete scenario of open inflation~\cite{Garcia-Bellido:2021idr} in which the causal horizon coincides with the boundary of a bubble wall separating our {\em flat}~FLRW universe from an open {\em empty}~space, therefore matching conditions require that both the scale factor and its derivative should be continuous at that hypersurface, i.e. $3H_{\rm in}^2 = \kappa\,\rho(a_{\rm in})$ and  $H_{\rm out}^2 = -k/a_{\rm in}^2$, which determines  $\sqrt{-k} = a_0H(a_0)$ today. We write the Hamiltonian constraint in conformal time (where primes denote derivatives w.r.t. $\tau=\sqrt{-k}\,\eta=a_0H_0\eta$) as~\footnote{Note that there is a difference w.r.t. Ref.~\cite{Garcia-Bellido:2021idr}, and we can now set $\OK=0$ inside the causal patch.}
\begin{eqnarray}
\left(\frac{a'}{a_0}\right)^2 = \OM\left(\frac{a}{a_0}\right) + 
 \frac{4\pi}{3}\left(\frac{a}{a_0}\right)^2 
\frac{\sinh(2\tau)}{(-k)^{3/2}V_c}  \,, \label{eq:Hamilton}
\end{eqnarray}
where the comoving volume (\ref{eq:Vc}) is now
\begin{equation}
    \label{eq:Vc2}
    (-k)^{3/2}V_c = \pi\left[\sinh(2\sqrt{-k}\,\eta_0) - 2\sqrt{-k}\,\eta_0\right]\,.
\end{equation}
We can then fix $\eta_0$ and determine the ratio between the causal horizon today and the curvature scale ($c=1$), 
\begin{equation}
    \label{eq:alpha}
    \alpha\,H(a_0)\,d_H(a_0)\equiv \frac{a_0\,\eta_0}{a_0/\sqrt{-k}} = \sqrt{-k}\,\eta_0\,.
\end{equation}
This ratio can be used to parametrize the different GREA scenarios and compare with $\Lambda$CDM, see Fig.~\ref{fig:dH}, where
\begin{equation}
    \label{eq:Horizon}
    H_0\,d_H(a) = \frac{2}{\sqrt\OM}\,a^{3/2}\cdot{}_2F_1\left[
    \frac{1}{2},\,\frac{1}{6},\,\frac{7}{6},
    \,\frac{\OM-1}{\OM}\,a^3\right]\,,
\end{equation}
is the expression for $d_H(a)$ in $\Lambda$CDM.

Solving the Hamiltonian constraint equation (\ref{eq:Hamilton}) with initial conditions set by the Cosmic Microwave Background, deep in the matter era, where $a_i(\tau)= \tau^2\,\OM/4$, with $\OM=0.31$, 
$\OK=0$  and $H_0=67.8$ km/s/Mpc~\cite{Planck:2018vyg}, we find an {\em effective} dark energy contribution $\OL\simeq 0.70$ today, for a narrow range of values of $\alpha\simeq1$, see Table \ref{tab:GREA}.


\begin{table}[h]
    \centering
    \begin{tabular}{| c | c | c | c | c | c | c |}
    \hline
     $\alpha$   &  $\OM$   &  $\OL$   &  $h_0$    &  $w_0$    &  $w_a$    &  $z_c$ \\ \hline
    \ 1.139 \ & \ 0.368 \ & \ 0.632 \ & \  0.786 \ & \ -0.908 \ & \  -0.315 \ & \ 0.931 \ \\ 
    \ 1.119 \ & \ 0.336 \ & \ 0.664 \ & \  0.749 \ & \ -0.937 \ & \  -0.331 \ & \ 0.840 \ \\ 
    \ 1.099 \ & \ 0.307 \ & \ 0.693 \ & \  0.714 \ & \ -0.966 \ & \  -0.345 \ & \ 0.757 \ \\ 
    \ 1.081 \ & \ 0.279 \ & \ 0.721 \ & \  0.682 \ & \  -0.996  \ &  \ -0.354 \  & \ 0.678 \ \\ 
    \ 1.062 \ & \ 0.254 \ & \ 0.746 \ & \  0.651 \ &  \ -1.028 \  & \  -0.365 \  &  \ 0.603 \ \\ 
    \ 1.043 \ & \ 0.231 \ & \ 0.769 \ & \  0.622 \ &  \ -1.060 \  & \  -0.367 \  &  \ 0.533 \ \\ 
    \hline\hline
    \  \ & \ 0.310 \ & \ 0.690 \ & \  0.678 \ &  \ -1.000 \  & \  -0.000 \  &  \ 0.645 \ \\ 
    \hline
    \end{tabular}
    \caption{The GREA values of the DM and DE content, the rate of expansion today, $H(0) = 100\,h_0$~km/s/Mpc, the effective EOS parameters ($w_0,\,w_a$) and the redshift $z_c$ of the correspon\-ding coasting point, for different values of the parameter $\alpha$. \JGB{From top to bottom, the values of $\alpha$ correspond to the (red, orange, yellow, green, blue, purple) colors in the rest of the figures.} We compare those values with the corresponding ones for $\Lambda$CDM (bottom row).}
    \label{tab:GREA}
\end{table}

\begin{figure}[t]
    \centering
    \includegraphics[width=0.494\linewidth]{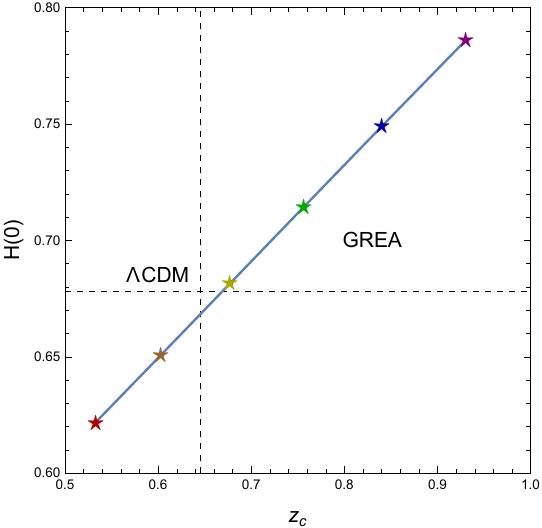}
    \includegraphics[width=0.494\linewidth]{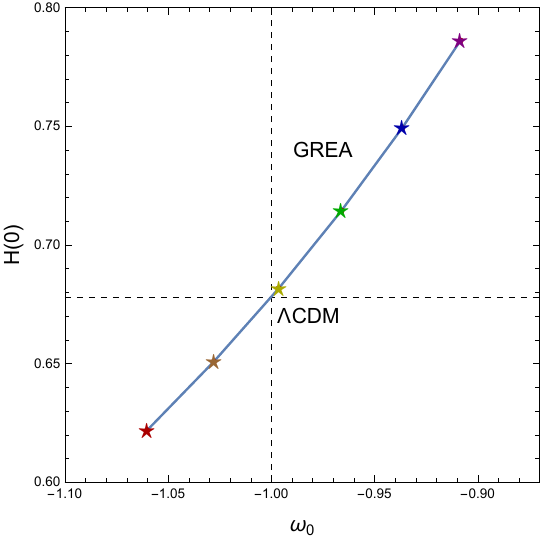}\\[3mm]
    \includegraphics[width=0.494\linewidth]{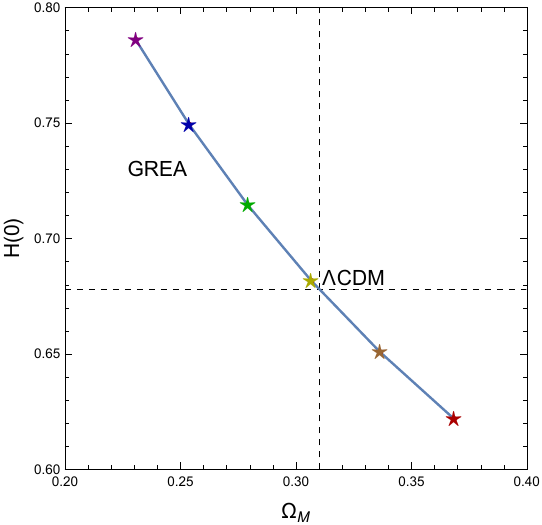}
    \includegraphics[width=0.494\linewidth]{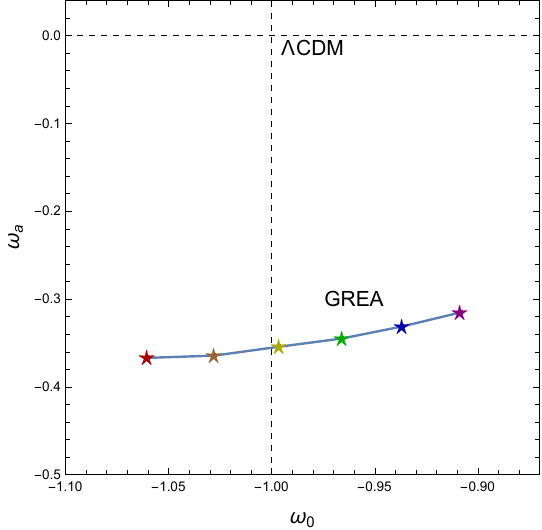}
    \caption{Different GREA scenarios in the plane of $\OM$, the coasting point $z_c$, the present rate of expansion $H(0)$ (in units of 100~km/s/Mpc) and the EOS parameters $(w_0,\,w_a)$. We also plot the $\Lambda$CDM values (black dashed lines). }
    \label{fig:Zc}
\end{figure}

It is important to emphasize that GREA has exactly the same number of free parameters as $\Lambda$CDM, i.e. ($\alpha,\,\OM,\,H_0$) versus ($\OL,\,\OM,\,H_0$), and nevertheless has very different dynamics, both at the present times and in the far future. Moreover, while in $\Lambda$CDM there is a coincidence (“why now") problem, i.e. why is $\OL\simeq\OM$ precisely today?, in the case of GREA the growth of the cosmic horizon during the matter era is responsible for the transition to a dominance of the entropic repulsive forces associated with the acceleration of the universe. The moment at which this happens depends on the actual value of the spatial boundary today in units of the causal horizon, as can be seen in Fig.~\ref{fig:CoastingPoint}a, where we show the coasting points for the different realizations of GREA. \JGB{In this scenario, the value of $\alpha$ depends on how far is our worldline from the bubble wall, as pictured in Fig.~1 of Ref.\cite{Garcia-Bellido:2021idr}. If the bubble wall would have been further away, the onset of acceleration would have happened later for the same value of $\OM$ inside our universe. However, in an open universe the largest volume is near the bubble wall, so it is expected that most observers will lie far from the center of the inflated bubble and near its edges. Therefore the finetuning associated with the coincidence problem is strongly alleviated.
Note also that different GREA scenarios populate the parameter space of the cosmological background in a way that can be easily distinguished from $\Lambda$CDM, see Fig.~\ref{fig:Zc}, with multiple panels, showing that of every 2D parameter space there is always a value of $\alpha$ that can accommodate a universe like that of $\Lambda$CDM, except for the value of $w_a$, which is always large and negative, in agreement with recent observations.}

\begin{figure}[t]
    \centering
    \includegraphics[width=\linewidth]{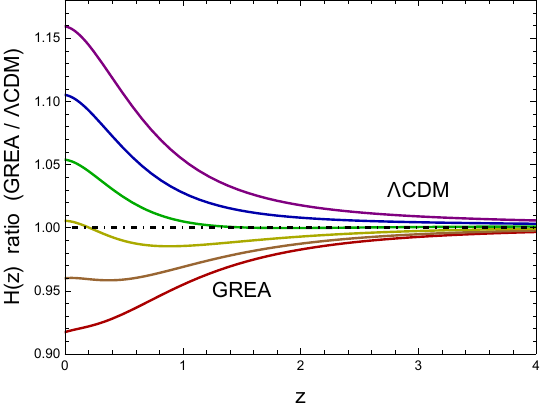}\\
    \includegraphics[width=\linewidth]{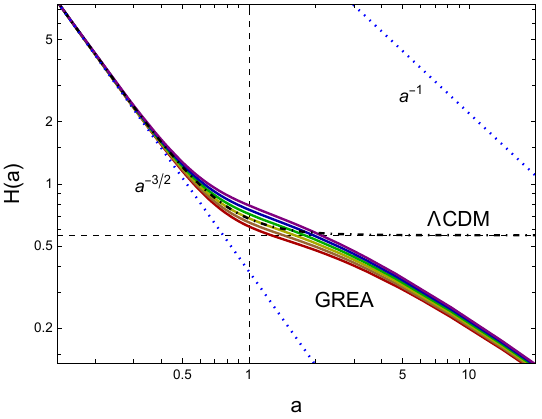}
    \caption{The upper panel shows the ratio of the evolution of the rate of expansion in GREA (color lines) versus $\Lambda$CDM (black dot-dashed line). The lower panel shows the actual evolution in GREA and $\Lambda$CDM, with the past (matter) and future (curvature) asymptotics (dotted blue lines).}
    \label{fig:RateExpansion}
\end{figure}

Furthermore, in GREA the value of the rate of expansion today is not fixed at the CMB, like in the case of $\Lambda$CDM. Here, depending on $\alpha$, the present rate of expansion, $H(0)$, could be larger or smaller than the $\Lambda$CDM value derived from the CMB, see Fig.~\ref{fig:RateExpansion}a and Table I. In fact, for certain values of the spatial curvature one can resolve in a natural way the so-called Hubble tension~\cite{DiValentino:2021izs}. 

Finally, while in $\Lambda$CDM, or its variant $w_0w_a\Lambda$CDM, the value of the equation of state of Dark Energy is given by $w(a) = w_0 + w_a(1-a)$, as an expansion around the present value, in GREA we have the whole function $w(a)$ to compare with observations, as can be seen in Fig.~\ref{fig:CoastingPoint}b. The evolution of the scale factor in GREA allows one to compute the {\em effective} equation of state of dark energy,
\begin{eqnarray}
1+w(a) &=& \frac{-d\ln}{d\ln a^3}\left[\frac{H^2(a)}{H_0^2} - 
\OM\!\left(\frac{a_0}{a}\right)^3\right]\nonumber \\[2mm]
&=& \frac{-d\ln}{d\ln a^3}\left[\frac{\sinh(2\tau)}{a^2(\tau)}\right]\,.
\end{eqnarray}
In the matter dominated era, $\tau\propto\sqrt{a}$, and therefore $1+w(a)\to1/2$ as $a\to a_i$, see Fig.~\ref{fig:CoastingPoint}b. In the far future, a combination of curvature and decaying entropy production gives a dependence $1+w(a)\to2/3$ as $a\to\infty$. The rate of expansion therefore will decay as curvature in the far future, diluting the universe, and ending in flat Minkowski space-time, see Fig.\ref{fig:RateExpansion}b. Note that this is very different from the standard $\Lambda$CDM scenario with the asymptote to empty de Sitter.

In fact, in GREA theory there is no freedom to chose ($\OM,\,H(0),\,w_0,\,w_a$) independently, as it happens in $w_0w_a\Lambda$CDM. The actual values of $(w_0,\,w_a)$ are precisely given by the theory, just like the present rate of expansion, $H(0) = 100\,h_0$ km/s/Mpc, once you fix the ratio $\alpha$, see Fig.~\ref{fig:Zc} for some examples. 
As a consequence, if the cosmological data favours values that are far away from $\Lambda$CDM $(w_0=-1,\,w_a=0)$, then we can test the whole theory by comparing those values with the corresponding predictions for both $H(0)$ and ($\OM,\,\OL$), as can be seen in Table~\ref{tab:GREA}.

\begin{figure}[t]
    \centering
    \includegraphics[width=\linewidth]{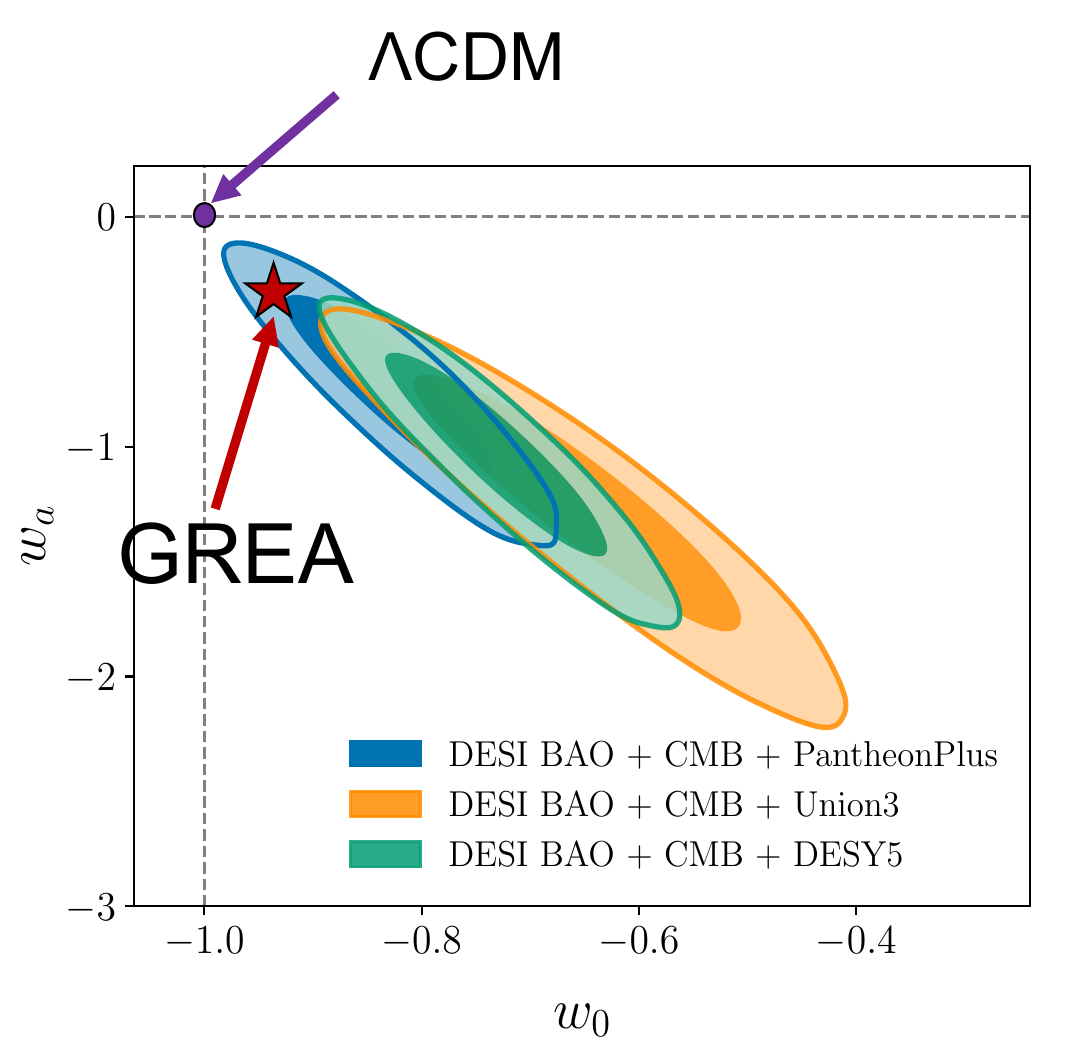}
    \caption{\JGB{The plot shows the parameter space $(w_0,\,w_a)$. Cosmological data seemed to prefer GREA from $\Lambda$CDM at the 2$\sigma$ level, already in 2021, see Ref.~\cite{Arjona:2021uxs}. Nowadays, the SN-Ia from DES Y5, and the BAO from DESI Y1, give compelling evidence that $\Lambda$CDM is excluded at close to 4$\sigma$, while GREA is within the 2$\sigma$ contours. 
    Figure adapted from Ref.~\cite{DESI:2024mwx}.}}
    \label{fig:H0w0wa}
\end{figure}

\JGB{We note in Fig.~\ref{fig:H0w0wa} that the GREA scenario is not in tension with the recent DESI-BAO~\cite{DESI:2024mwx} + DES-SNIa~\cite{DES:2024tys} data, although a proper analysis, including the background and linear perturbations of GREA, has to be performed with the full data set (CMB, SNIa, LSS, BAO, RSD, ISW, $H_0$, etc.). In fact, the amount of information that we can extract from present observations is larger than what can be inferred from their projection on to the plane ($w_0,\,w_a$), and future data will be even more constraining at different redshifts.}

\section{IV. Linear perturbation theory in GREA}

So far the description of GREA has been done at the level of the background evolution, while we have ignored the small matter fluctuations that give rise to galaxies and the large scale structure (LSS) of the universe. In order to understand the effect of GREA on the growth of matter perturbation we need to develop a linear perturbation theory in the accelerating background of GREA, which will drive a rate of growth that may be distinguishable in principle from that of $\Lambda$CDM. Furthermore, in order to compare with observations, this linear growth has to be properly characterized as a function of the free parameters of the theory.

The equations driving the evolution of matter fluctuations can be derived from the covariant conservation of the total energy-momentum tensor~(\ref{eq:Tmunu}). The background metric in conformal time is that of a curved FLRW with a Newtonian potential $\Phi$ and curvature fluctuation $\Psi$,
\begin{equation}
	\label{eq:FLRW}
	ds^2 = a^2(\eta)\left[- (1+2\Phi)\,d\eta^2 + (1-2\Psi)\gamma_{ij}dx^i dx^j\right]\,.
\end{equation}
In the absence of shear and vorticity (i.e. for a 
spatially-symmetric matter energy-momentum tensor), the Newtonian potential and curvature fluctuation satisfy $\Phi = \Psi$. Then the Bardeen equation~\cite{Bardeen:1980kt} for the gauge invariant curvature fluctuation in flat space ($K=0$) with a cosmological constant {\em and} entropic pressure can be written, in terms of a general equation of state $w=\sum p_i/\sum \rho_i$, as 
\begin{equation}
	\label{eq:Bardeen}
	\Phi''(\eta) + 3{\cal H}\Phi'(\eta) - 3w{\cal H}^2\Phi = c_s^2\,\nabla^2\Phi
    \,,
\end{equation}
where the Poisson equation
\begin{equation}
	\label{eq:Poisson}
	\nabla^2\Phi = 4\pi G\rho\,a^2\,\delta
\end{equation}
relates the curvature fluctuation $\Phi$ to the gauge-invariant density contrast, $\delta$.
Expanding Bardeen's equation (\ref{eq:Bardeen}) for a matter fluid with negligible speed of sound, $c_s^2\simeq0$, we find
\begin{equation}
	\label{eq:Bardeen2}
	\Phi''(\eta) + 3{\cal H}\Phi'(\eta) + (\Lambda a^2 - 8\pi G\,a^2 p_S)\Phi(\eta) = 0
    \,,
\end{equation}
with $p_S = - \rho_H$ is the entropic pressure (\ref{eq:Hamilton}). We can then rewrite the Bardeen equation, using the Poisson equation, $\Phi \propto \delta/a$, and the identity
$${\cal H}'+2{\cal H}^2 = \frac{3}{2}H_0^2\frac{\OM}{a} + \Lambda a^2 - 8\pi G\,a^2 p_S,$$
to finally write
\begin{equation}
	\delta''(\tau) + \frac{a'}{a}\,\delta'(\tau) - \frac{3}{2}\,\frac{\OM}{a(\tau)}\,\delta(\tau) = 0\,,
 \label{eq:master}
\end{equation}
where $\tau=a_0H_0\eta$, and the time evolution is determined by eq.(\ref{eq:Hamilton}). We can now integrate with initial condition deep in the matter era $\delta_i(\tau)\propto a_i(\tau) \propto \tau^2$. Once the solution is found for the density contrast, we can derive the time evolution of the Newtonian potential $\Phi$ via the Poisson equation for matter fluctuations~(\ref{eq:Poisson}).

\begin{figure}[t]
    \centering
    \includegraphics[width=\linewidth]{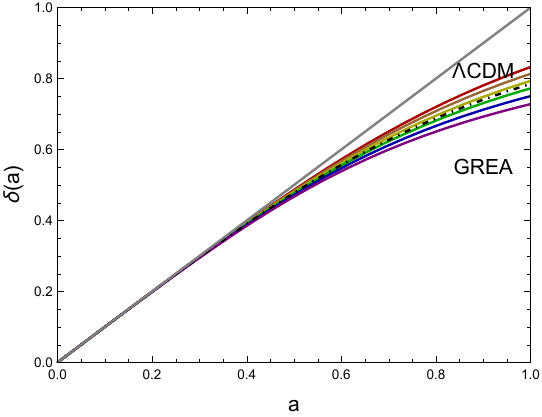}\\
    \vspace{5mm}
    \includegraphics[width=\linewidth]{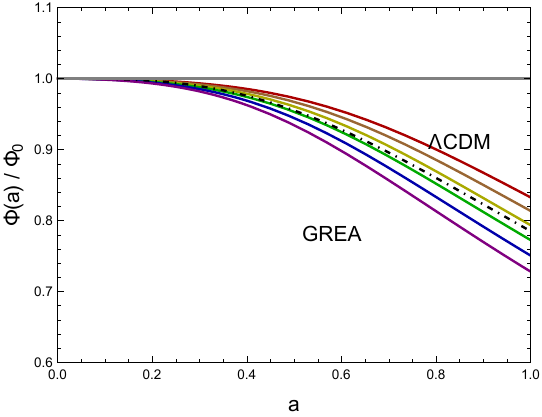}
    \caption{We show the damped evolution of the density contrast (upper panel) and the Newtonian potential (lower panel), as a function of the scale factor for $\Lambda$CDM (black dot-dashed line) and GREA (colors correspond to those of Fig.~\ref{fig:CoastingPoint}). The gray line corresponds to the growth in an open matter dominated cosmology. It is clear that both $\Lambda$CDM and GREA perturbations deviate significantly from a matter-only scenario after $a\sim0.2$ or $z\sim4$.}
    \label{fig:delta}
\end{figure}

For flat $\Lambda$CDM, the solution to eq.~(\ref{eq:master}) can be written explicitly in terms of Hypergeometric functions,
\begin{eqnarray}
    \label{eq:deltaLCDM}
    \delta(a) &=& \delta_0 \,a \cdot {}_2F_1\left[1,\,
    \frac{1}{3},\,\frac{11}{6},\,\frac{\OM-1}{\OM}\,a^3\right]\,, \\[2mm]
    \tau(a) &=& \frac{2\sqrt{a}}{\sqrt\OM}\, \cdot {}_2F_1\left[
    \frac{1}{2},\,\frac{1}{6},\,\frac{7}{6},
    \,\frac{\OM-1}{\OM}\,a^3\right]\,, \label{eq:tau}
    \\[2mm]
    \Phi(a) &=& \Phi_0\, \cdot {}_2F_1\left[1,\,
    \frac{1}{3},\,\frac{11}{6},\,\frac{\OM-1}{\OM}\,a^3\right]\,.
    \label{eq:Phi}
\end{eqnarray}
In the case of GREA, we must solve first the background evolution (\ref{eq:Hamilton}), and then the linear perturbation eq.~(\ref{eq:master}). We compare in Fig.~\ref{fig:delta} the time evolution of both the density contrast
$\delta(a)$ and the Newtonian potential $\Phi(a)$
for $\Lambda$CDM and GREA for the values of $\alpha$ in Table~\ref{tab:GREA}.

\begin{figure}[t]
    \centering
    \includegraphics[width=\linewidth]{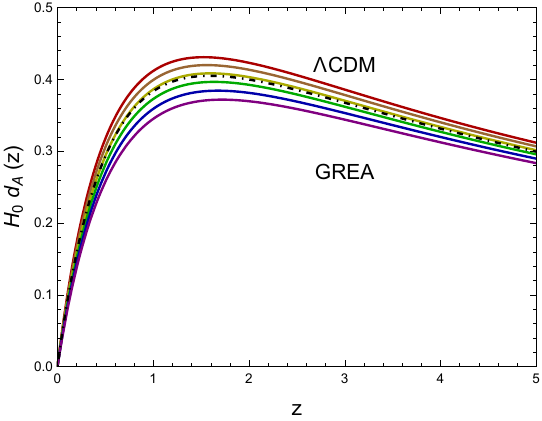}
    \caption{We show the angular diameter distance for $\Lambda$CDM (black dot-dashed line) and GREA (colors correspond to those of Fig.~\ref{fig:CoastingPoint}).}
    \label{fig:dA}
\end{figure}

\subsection{The growth function and index}

\begin{figure}[t]
    \centering
    \includegraphics[width=\linewidth]{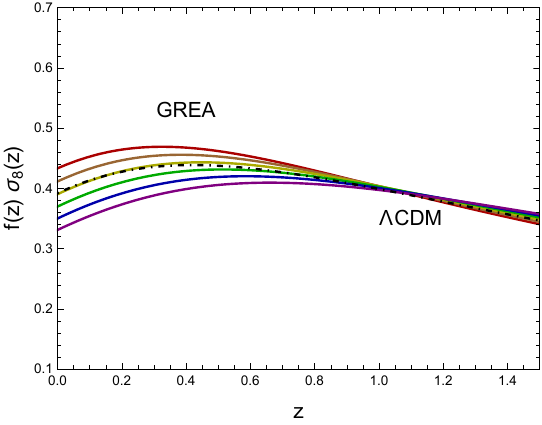}\\[2mm]
    \includegraphics[width=\linewidth]{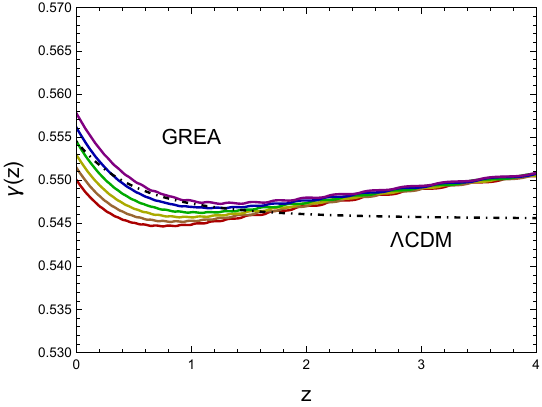}
    \caption{We show the growth function $f(z)\,\sigma_8(z)$ and growth index $\gamma(z)$ for $\Lambda$CDM (black dot-dashed line) and GREA (colors correspond to those of Fig.~\ref{fig:CoastingPoint}).}
    \label{fig:gamma}
\end{figure}

In Large Scale Structure, the redshift-space distortions in the two-point correlation function $\xi(r)$ depends on a particular combination of the density contrast that appears in the velocity perturbation known as the growth function~\cite{Amendola:2016saw}, $f(a) = d\ln\delta(a)/d\ln a$, where $\delta(a)$ is the linear matter perturbation. In the case of $\Lambda$CDM the growth function has a compact expression~\cite{Buenobelloso:2011sja},
\begin{equation}
	f(a) = \OM^{1/2}(a)\frac{P_{-1/6}^{-5/6}\Big[\OM^{-1/2}(a)\Big]}
    {P_{1/6}^{-5/6}\Big[\OM^{-1/2}(a)\Big]} \ \equiv \ \OM^\gamma(a)\,,
 \label{eq:growth}
\end{equation}
where $\OM(a) = \OM/a^3H^2(a)$ and $P_n^m(z)$ are the associated Legendre polynomials. The growth index $\gamma$ is defined as the power of $\OM(a)$ in the growth function,
\begin{equation}
	\gamma(a) = \frac{1}{2} + 
    \frac{1}{\ln\OM(a)}\ln\left[\frac{P_{-1/6}^{-5/6}
    \Big[\OM^{-1/2}(a)\Big]}
    {P_{1/6}^{-5/6}\Big[\OM^{-1/2}(a)\Big]}\right] \,.
 \label{eq:growthindex}
\end{equation}
We have shown in Fig.~\ref{fig:gamma} the growth index for $\Lambda$CDM and GREA for different values of $\OK$ as a function of redshift. Future surveys should be able to distinguish between these two scenarios.

\subsection{The BAO scale and the $S_8$ tension}

One of the most important physical quantities that one can measure with deep galaxy surveys is the BAO scale, corresponding to the size of the sonic horizon as a function of redshift. The angular BAO scale depends on both the sonic horizon at the baryon drag epoch, $\chi_s(z_{\rm drag})= 110$ Mpc/h, and the angular diameter distance, while the radial BAO scale depends also on the rate of expansion,
\begin{eqnarray}
	\theta_{\rm BAO}(z) &=& \frac{\chi_s(z_{\rm drag})}{d_A(z)}\,, \\
	r_{\rm BAO}(z) &=& \frac{H(z)}{c}\,\chi_s(z_{\rm drag})\,. 
 \label{eq:BAO}
\end{eqnarray}

The angular diameter distance for flat $\Lambda$CDM is given explicitly in terms of $\tau(a)$ of eq.~(\ref{eq:tau})
\begin{equation}
	H_0\,d_A(z) = \frac{\tau(0) - \tau(z)}{1+z}\,.
    \label{eq:dA}
\end{equation}
We show in Fig.~\ref{fig:dA} the angular diameter distance as a function of redshift. 

The luminosity distance needed to make connection with the SN-Ia data is simply related to the angular diameter distance via the Etherington relation, $$d_L(z) = (1+z)^2\,d_A(z)\,.$$

\begin{figure}[t]
    \centering
    \includegraphics[width=\linewidth]{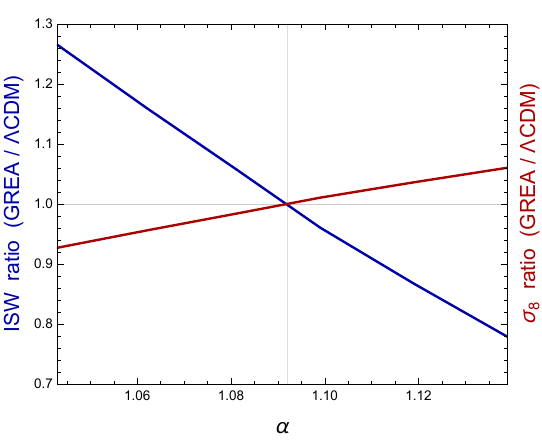}\\
    \caption{We show the ratio of the amplitude of fluctuations (red) at 8 Mpc/h and the ISW effect (blue) between GREA and $\Lambda$CDM, as a function of $\alpha$. There is a value for which both ratios are one, for $\alpha=1.092$.}
    \label{fig:sigma8}
\end{figure}

Another cosmological observable which is derived from weak lensing is the magnitude of the fluctuations today on a scale of 8 Mpc/h. For $\Lambda$CDM and GREA we have different predictions and one can write the ratio as a function of $\alpha$, as shown in Fig.~\ref{fig:sigma8}. For those values of $\alpha$ that resolve the $H_0$ tension, predicting today $H(0) = 73\pm1$ km/s/Mpc~\cite{DiValentino:2021izs}, the $\sigma_8$ value predicted by GREA is around 4\% lower than that of $\Lambda$CDM, resolving also the $S_8$ tension~\cite{DiValentino:2021izs}.

\JGB{We should be cautious here, since in order to claim a resolution of the so-called $H_0$ and $S_8$ tensions within GREA requires a full CMB, weak lensing and Cepheid-TRGB analysis. In fact, recent claims of resolution of each of these tensions have been discussed recently and it could be that they are associated with unknown systematics yet to be understood.}

\subsection{The ISW effect}

An important observable that can be obtained from the CMB maps is the Integrated Sachs-Wolfe (ISW) effect, which arises because of the decrease in the Newtonian potential $\Phi$ due to the presence of an accelerating expansion, as photons travel since decoupling and feel the redshift induced by the expansion of the Universe. This corresponds to a temperature decrement given by
\begin{equation}
	\frac{\delta T}{T}= 2\int_0^{\eta_0} d\eta\,\Phi'(\eta_0 - \eta)\,.
    \label{eq:ISW}
\end{equation}
The equation for $\Phi(\eta)$ can be derived in flat $\Lambda$CDM as $\Phi''(\tau) + 3(a'/a)\,\Phi'(\tau) + \Lambda\,a^2 \Phi(\tau)= 0$, see eq.~(\ref{eq:Bardeen2}). 

For GREA the Newtonian potential evolves according to the density contrast, see Fig.~\ref{fig:delta}b. One can integrate along the line of sight from the last scattering surface, see eq.~(\ref{eq:ISW}), and compare with the predicted value in $\Lambda$CDM. We show the ratio in Fig.~\ref{fig:sigma8}.
While the effect of GREA on $\sigma_8$ can be at most a few percent, in the case of the ISW effect, the correction can reach 20 or 30\%.

Note that we have studied here only the homogenous ISW effect, due to the global expansion of the universe. There is a much more pronounced ISW effect towards deep voids in the cosmic web, where photons traversing large voids get an extra decrement in the ISW temperature contrast due to the growth of the Newtonian potential as the void becomes emptier. Such an effect is known to be significantly larger and cannot be accounted for by the evolution of the LSS around large voids in $\Lambda$CDM~\cite{DES:2018nlb}. However, GREA can account for this extra acceleration due to the production of entropy associated with the formation of the cosmic web and the growth of voids as a consequence of gravitational collapse of large structures like filaments, sheets and superclusters~\cite{Garcia-Bellido:2024tip}. This effect is not computed here and may be relevant for the comparison with observations in the future.

\section{V. Conclusions}

General Relativistic Entropic Acceleration Theory is a covariant framework that can explain the present acceleration of the universe from the entropic force associated with the growth of the cosmological horizon, without the need to introduce a {\em fundamental} cosmological constant, i.e. $\Lambda=0$ always. 
The cosmic entropic acceleration associated with the homogeneous expansion of the cosmological causal horizon can be described in terms of just one parameter $\alpha$, the ratio between the horizon distance and the spatial curvature.  

\JGB{The entropic acceleration due to the causal horizon starts to be important only recently, thanks to the $\sinh(2\tau)$ dependence in Eq.~(\ref{eq:Hamilton}), alleviating the coincidence problem. It is the fact that the universe is old and big that GREA started to dominate the expansion of the universe only recently. In the past, such a term was exponentially negligible, for ratios $\alpha\simeq1$. One only has to live near the edge of the bubble, where most of the volume is. For larger horizon distances the acceleration would have started later for the same amount of matter.}

The recent evidence for deviations from the $\Lambda$CDM paradigm seems to agree remarkably well with the predictions of GREA at the homogeneous background level. In particular, it resolves the Hubble tension by shifting the coasting point $z_c$ to higher redshifts and extending the period of accelerated expansion with respect to $\Lambda$CDM. Nevertheless, this entropic acceleration doesn't last for ever and eventually the GREA entropic term is diluted just like matter, so that the universe ends in a flat Minkowski space-time, instead of de Sitter. 

We find that some background observables in GREA, like $\OM$ and the effective equation of state parameter today $w_0$, remain very close to those of $\Lambda$CDM, for values of $\alpha\simeq1$, see Fig.~\ref{fig:CoastingPoint}. 
In fact, we find a dark energy parameter today, $\OL$, consistent with observations, which  gives an {\em effective} $\Lambda$ that is 120 orders of magnitude below the Planck scale, without finetuning. Meanwhile, for the same values of $\alpha$, the EOS slope $w_a$ is found to be significantly large and negative, and closer to the observed values today, without introducing a new parameter to explain the data.

We have also studied the linear growth of matter perturbations in the context of the homogeneous expanding background driven by the growth of entropy from the cosmological horizon. We find that the rate of growth of matter fluctuations in GREA slows down due to the accelerated expansion and alleviates the $\sigma_8$ tension. We further study the redshift-space distortions induced by the growth function of matter fluctuations, which enters the two-point correlation function of galaxies measured in LSS surveys. We find that the corresponding ISW effect is significantly larger than in $\Lambda$CDM, but not enough to explain the observed anomalous ISW effect.

The main effect of GREA on the formation of structure is the accelerated expansion due to the rapid growth of entropy of the cosmological horizon during the matter era. However, there are also local accelerations due to the growth of entropy associated with the growth of entropy associated with the gravitational collapse of large scales structures in the cosmic web, which is responsible for emptying the large voids and can induce an additional ananomalous ISW effect. These {\em inhomogeneous} local accelerations have not been discussed here and will be the subject of a future publication.

In summary, it is encouraging that most of the tensions and anomalies of the standard model of cosmology are resolved, or at least alleviated, by the inclusion of this transient period of acceleration of the universe based on known fundamental physics.  
We leave for a future publication the detailed comparison of the predictions of GREA theory with the present cosmological data from deep and wide galaxy surveys like DES, DESI and Euclid.

\section{Acknowledgements}

The author thanks Julien Lesgourgues for constructive criticisms on a preliminary draft that has helped improve the paper. 
He also acknowledges support from the Spanish Research Project PID2021-123012NB-C43 [MICINN-FEDER], and the Centro de Excelencia Severo Ochoa Program CEX2020-001007-S at IFT. 

\bibliography{main}

\end{document}